\documentclass[twocolumn,showpacs,preprintnumbers,amsmath,amssymb, floatfix]{revtex4}

\usepackage{graphicx}
\usepackage{dcolumn}
\usepackage{bm}
\usepackage{mathrsfs}
\usepackage{epsfig, subfigure}
\usepackage{multirow}

\newcommand{\Lag}{\mathscr{L}}
\newcommand{\Tr}{{\rm Tr}}
\newcommand{\td}{{\rm d}}

\begin{document}

\preprint{}

\title{Anomalous Quartic Gauge Couplings from Six Quark Production}

\author{Erik Schmidt}
\author{Michael Beyer}
\author{Henning Schr\"oder}%
\affiliation{%
Institute of Physics, University of Rostock, 18051 Rostock, Germany\\
}%

\date{\today}

\begin{abstract}

The absence of a light Higgs boson causes vector boson couplings to
become strong at 1 TeV. A general framework for a systematic and
consistent treatment is provided by effective theories of electroweak
symmetry breaking.  Already in next-to-leading order
there appear quartic gauge couplings that go beyond the
standard model and are hence called anomalous.  We investigate
intermediate three gauge boson states $W^+W^- Z$ and $ZZZ$ occurring
in six quark production in electron positron collisions under the
conditions of the International Linear Collider. We perform a
sensitivity analysis of the relevant anomalous quartic gauge couplings
presenting their expected limits.

\end{abstract}



\maketitle
\sloppy
\section{Introduction}

One important objective of future accelerator experiments is to unveil
the mechanism behind electroweak symmetry breaking (EWSB).  In this
respect generic effective field theories are a proper choice for a
systematic and consistent treatment.  These theories arise in a
natural way, if no light Higgs is assumed to be present in the
particle spectrum \cite{Susskind:1978ms}.  Including all operators
obeying the underlying symmetries, these effective theories are
non-renormalizable in a narrow sense. At each order of perturbation
theory more counterterms have to be introduced in order to render the
whole theory finite. These so called anomalous couplings parameterize
the effects of new physics in a generic, model independent way.  Since
the construction of the theory starts from the known low energy
behavior of the electroweak interaction, the way of describing EWSB is
called the bottom-up approach, see
e.g. ref. \cite{Kilian:2003pc}. Alternatively, the top-down approach
constitutes the whole particle spectrum and symmetries of the theory
at high energies from the beginning by constructing renormalizable
Lagrangians.

In the following we consider the interaction between four bosons, in
particular production of intermediate three boson states. The relevant
CP conserving operators arising in next-to-leading-order are the two
$SU(2)_V$-conserving \cite{Longhitano:1980tm}
\begin{align}
\Lag_4 =& \frac{\alpha_4}{16\pi^2}\Tr(V_\mu V_\nu)\Tr(V^\mu V^\nu)\label{eq:L4}\\
\Lag_5 =& \frac{\alpha_5}{16\pi^2}\Tr(V^\mu V_\mu)^2,\label{eq:L5}
\end{align}
with $V_\mu = \Sigma D_\mu\Sigma^\dagger$, and $D_\mu$ being the $SU(2)\times U(1)$-gauge-covariant
derivative.  The field $\Sigma$ parameterizes the Goldstone sector and provides a generally 
nonlinear realization of the electroweak gauge symmetry
\cite{Kilian:2003pc}. In unitary gauge $\Sigma\equiv 1$ the operator
$V_\mu$ reduces to \begin{equation}V_\mu =
-i\frac{g_W}{\sqrt{2}}(W^+_\mu \tau^+ + W^-_\mu\tau^-)-ig_Z
Z_\mu\tau^3, \end{equation} where $\tau^\pm$ and $\tau^3$ are the
generators of weak isospin group $SU(2)$.  Three
further operators contributing to a generic four-point-interaction
are
\begin{align}
\Lag_6 =& \frac{\alpha_6}{16\pi^2}\Tr[V_\mu V_\nu]\Tr[TV^\mu]\Tr[TV^\nu]\\
\Lag_7 =& \frac{\alpha_7}{16\pi^2}\Tr[V_\mu V^\mu]\Tr[TV_\nu]\Tr[TV^\nu]\\
\Lag_{10} =& \frac{\alpha_{10}}{32\pi^2}[\Tr(TV_\mu)\Tr(TV^\mu) ]^2,
\end{align}
which, in contrast to Eqs. (\ref{eq:L4}, \ref{eq:L5}), violate $SU(2)$ via the appearance of 
the operator $T=\Sigma\tau^3\Sigma^\dagger$.

These interaction terms are evaluated in three-boson-production via the processes
\begin{eqnarray}
\begin{split}\label{proc}
 e^+ e^- &\longrightarrow (W^+W^-Z)&\longrightarrow q\bar q  q\bar q q\bar q\\
 e^+ e^- &\longrightarrow (ZZZ) & \longrightarrow q\bar q  q\bar q q\bar q.
\end{split}
\end{eqnarray}
Only the sum of $SU(2)$-conserving and -violating operators
contributes to each of these processes. Hence, there will be no direct
sensitivity to $SU(2)$-violation.  This question can be accessed by
complementary analysis, such as W-scattering, see e.~g.
\cite{Krstonosic:2005qp,Beyer:2006hx}.  The coupling parameters can
be mapped to masses of resonant particles that may be indirectly accessible
in the range of future colliders \cite{Beyer:2006hx, Kilian:2005bz}.

The results presented here are based on previous analyses
\cite{Beyer:2006hx}. In this paper we now extend this work in
two directions: First, we consider the influence of the polarization
states of the bosons. This is particularly interesting since the
longitudinal polarized components directly correspond to the would-be
Goldstone bosons associated with the symmetry breaking
pattern. Second, the restriction of considering on-shell bosons in the
final state is abandoned in favor of investigating the complete set of
six quark final states, corresponding to the pure hadronic decay of
the intermediate bosons, which brings the overall analysis closer to
the future experimental situation.  

\begin{figure}[htb]
\caption{\label{fig:hel_ang}Reconstructed differential cross sections normalized to the SM cross section. }
\epsfig{file=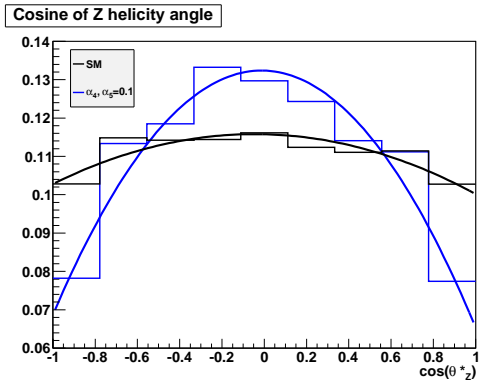,width=0.4\textwidth}
\end{figure}

For a motivation of how the helicity angle may improve sensitivity we
show Fig.~\ref{fig:hel_ang}, presenting the reconstructed helicity
distribution in dependence of the anomalous coupling parameters.
We have chosen $\alpha_4=\alpha_5=0.1$ and compare the helicity
distributions implied by taking into account the anomalous couplings to
the one of the standard model. The distribution for $\cos\theta^*_Z$
are given after reconstruction as explained in the next sections.

To guide the eye the histogram has been fitted by a
simple second order polynomial.  From inspection of
Fig.~\ref{fig:hel_ang} it is obvious that the appearance of anomalous
couplings change the shape of the distribution from a rather flat one
to a more pronounced distribution.  Although the helicity state of the
intermediate boson is not reconstructed this way it is possible to
take into account the effects of the intermediate state of
polarization.
 
The helicity angle is given with respect to the rest system of the
decaying boson, in this case the $Z$ boson in $WWZ$ intermediate
states. To this end the co-ordinate system is properly boosted
utilizing the reconstructed momenta (subject to kinematical fit).
 
The curves in Fig.~\ref{fig:hel_ang} are normalized to the respective
total number of events to separate effects on total cross section from
variations in the distribution itself.

\section{Analysis Layout}

\begin{table}[t]
\caption{\label{tab:6qstates} Generated six quark production processes at 
$\sqrt{s}=1$ TeV considered in the analysis. (Cross sections including numerical errors are given).}
\begin{tabular}{ccc}
\hline
\hline
out state & \parbox{2.5cm}{intermediate state} & $\sigma_{\rm tot}\ [{\rm fb}]$ \\[0.5ex]
\hline
$d\bar d d\bar d d\bar d$ & $ZZZ$ & $(1.886 \pm 0.011)\cdot 10^{-2}$\\
$d\bar d\bar dud\bar u$ & $WWZ$ & $0.228\pm0.005$\\
$d\bar dd\bar ds\bar s$ & $ZZZ$ & $(5.51\pm 0.05)\cdot 10^{-2}$\\
$d\bar dd\bar dc\bar c$ & $ZZZ$ & $(8.80\pm 0.07)\cdot 10^{-2}$\\
$d\bar dd\bar db\bar b$ & $ZZZ$ & $(7.24\pm0.04)\cdot 10^{-2}$\\
$d\bar dd\bar u\bar sc$ & $WWZ$ & $0,1442\pm 0.0006$\\
$d\bar ds\bar c\bar du$ & $WWZ$ & $0.1449\pm 0.0009$\\
$d\bar u\bar duu\bar u$ & $WWZ$ & $0.221\pm 0.005$ \\
$d\bar u\bar dus\bar s$ & $WWZ$ & $0.277\pm 0.005$\\
$d\bar u\bar duc\bar c$ & $WWZ$ & $0.339\pm 0.007$\\
$d\bar u\bar dub\bar b$ & $WWZ$, $tt$ & $17.72\pm 0.04$\\
\hline
\hline
$d\bar ds\bar ss\bar s$ & $ZZZ$ & $(5.95\pm0.11)\cdot 10^{-2}$\\
$d\bar ds\bar c\bar sc$ & $WWZ$ & $0.300\pm 0.008$\\
$d\bar ds\bar sb\bar b$ & $ZZZ$ & $0.1471\pm0.0016$\\
$d\bar dc\bar cc\bar c$ & $ZZZ$ & $0.1157\pm 0.0023$\\
$d\bar dc\bar cb\bar b$ & $ZZZ$ & $0.2231\pm 0.0027$\\
$d\bar db\bar bb\bar b$ & $ZZZ$ & $(8.10\pm 0.06)\cdot 10^{-2}$\\
$d\bar uu\bar u\bar sc$ & $WWZ$ & $0.1317\pm 0.0005$\\
$d\bar us\bar s\bar sc$ & $WWZ$ & $0.1440\pm 0.0009$\\
$d\bar u\bar scb\bar b$ & $WWZ$, $tt$ & $17.548\pm 0.029$\\
$d\bar u\bar scc\bar c$ & $WWZ$ & $0.1316 \pm 0.0006$\\
$u\bar u\bar dus\bar c$ & $WWZ$ & $0.1317 \pm 0.0005$\\
\hline
\hline
$s\bar s\bar dus\bar c$ & $WWZ$ & $0.1448\pm 0.0005$\\
$s\bar c\bar duc\bar c$ & $WWZ$ & $0.1324\pm 0.0007$\\
$s\bar c\bar dub\bar b$ & $WWZ$, $tt$ & $17.60\pm 0.04$\\
$u\bar uu\bar uu\bar u$ & $ZZZ$ & $(4.53\pm 0.03)\cdot 10^{-2}$\\
$u\bar uu\bar us\bar s$ & $ZZZ$ & $0.1152\pm 0.0012$\\
$u\bar uu\bar uc\bar c$ & $ZZZ$ & $0.1374\pm 0.0015$\\
$u\bar uu\bar ub\bar b$ & $ZZZ$ & $0.1430\pm 0.0009$\\
$u\bar us\bar ss\bar s$ & $ZZZ$ & $(8.54\pm 0.12)\cdot 10^{-2}$\\
$u\bar us\bar c\bar sc$ & $WWZ$ & $0.356\pm 0.012$\\
$u\bar us\bar sb\bar b$ & $ZZZ$ & $0.2206\pm 0.0021$\\
$u\bar uc\bar cc\bar c$ & $ZZZ$ & $0.131\pm 0.004$\\
\hline
\hline
$u\bar uc\bar cb\bar b$ & $ZZZ$ & $0.281\pm 0.006$\\
$u\bar ub\bar bb\bar b$ & $ZZZ$ & $0.1215\pm 0.0007$\\
$s\bar ss\bar ss\bar s$ & $ZZZ$ & $(1.926\pm 0.014)\cdot 10^{-2}$\\
$s\bar ss\bar sc\bar c$ & $ZZZ$ & $0.2269\pm 0.0022$\\
$s\bar ss\bar sb\bar b$ & $ZZZ$ & $(7.26\pm 0.05)\cdot 10^{-2}$\\
$s\bar sc\bar cc\bar c$ & $ZZZ$ & $0.234\pm 0.006$\\
$s\bar c\bar scb\bar b$ & $WWZ$, $tt$ & $17.77\pm 0.04$\\
$s\bar sb\bar bb\bar b$ & $ZZZ$ & $(8.11\pm 0.06)\cdot 10^{-2}$\\
$c\bar cc\bar cc\bar c$ & $ZZZ$ & $(4.55\pm 0.04)\cdot 10^{-2}$\\
$c\bar cc\bar cb\bar b$ & $ZZZ$ & $0.1416\pm 0.0009$\\
$c\bar cb\bar bb\bar b$ & $ZZZ$ & $0.1215\pm 0.0009$\\
$b\bar bb\bar bb\bar b$ & $ZZZ$ & $(2.638 \pm 0.011)\cdot 10^{-2}$\\
\hline
\hline
\end{tabular}
\end{table}

We analyze intermediate three-boson-production at a center-of-mass
energy of $\sqrt{s} = 1\ {\rm TeV}$. The total integrated luminosity
is assumed to be $\int\mathcal{L} = 1000\ {\rm fb}^{-1}$.  An earlier
analysis has shown \cite{Krstonosic:2005qp} that the sensitivity
increases remarkably by using polarized initial $e^+e^-$-states.  We
therefore focus on the situation with 80 \% right handed polarization
of electrons and 60 \% left handed polarization of positrons that has
been the most favorable scenario. The degree of polarization has been 
carried over from our earlier analysis.

We consider all the six quark final states listed in Tab.~\ref{tab:6qstates}.

Each of the flavor final states is equivalent by a large number of Feynman
graphs. They contain the signal events as well as a huge amount of
background arising from (a few thousand) standard model
graphs with the same in and out states but without any quartic
gauge couplings. (Note, however, that the actual generator algorithm
 differs from evaluating Feynman graphs, but using technique 
 described closer in \cite{Moretti:2001zz})

The dominating part of this background stems from direct top
production, via
\begin{equation} 
e^+ e^- \longrightarrow t\bar t \longrightarrow
q\bar q q\bar q b\bar b,
\end{equation} 
which manifests itself as a six-jet-topology.  They are indicated as
$tt$ in Tab.~\ref{tab:6qstates}.

An additional type of background is related to misidentification. In case of
the $ZZZ$ intermediate states, also the contamination of reconstructed
$Z$s with misidentified $W$s has to be taken into account. Particle
misidentification amounts to 8\% of all reconstructed $Z$s.  The
converse case ($Z$ identified as $W$ in reconstructed $WWZ$ events) is
less important, since the $ZZZ$ production is largely suppressed
compared to $WWZ$ production.

The generation of events is achieved by using {\sc Whizard}
\cite{Kilian:2001qz} and {\sc Pythia} \cite{Sjostrand:2003wg} is utilized for
fragmentation to hadronic final states.  Initial state radiation (ISR)
has been taken into account.  In order to reduce statistical
fluctuations, the number of analyzed events is increased by a factor
of hundred of the expected number (assuming a luminosity of
$\mathcal{L}= 1000$ fb${}^{-1}$ at the ILC) and finally renormalized.
This procedure leads to a more reliable figure of merit for the
sensitivity.  Subsequent detector simulation is done with the fast
simulation tool {\sc Simdet} \cite{Pohl:2002vk}.

\subsection{Fit Method}

We use a $\chi^2$-fit to extract information on anomalous
couplings from the distribution of kinematical variables
\begin{equation}\label{eq:chi2}
\chi^2 =\sum_{ijk\ell mn} \frac{\left(N_{ijk\ell mn}^{\rm
th}(\alpha_4, \alpha_5) - N^{\rm exp}_{ijk\ell
mn}\right)^2}{\sigma_{ijk\ell mn}^2}.
\end{equation}
The indices $ik\ell mn$ stand for the discretized phase space
variables, which are six in number, and are used to describe the event.  They
will be dropped in the following formulas for convenience.
Confidence levels for $\alpha_4$ and $\alpha_5$ are calculated with
{\sc Minuit} \cite{James:1975dr}.  The kinematical variables, the
distributions of which are sensitive observables to anomalous quartic gauge couplings and are
therefore used in the analysis, are
\begin{itemize}
\item the invariant mass $M_{WZ}=\sqrt{(p_{W_1} + p_Z)^2}$ of the $W_1$-$Z$-subsystem
\item the invariant mass $M_{WW}=\sqrt{(p_{W_1} + p_{W_2})^2 }$ of the $W_1$-$W_2$-subsystem
\item the cosine $\cos\theta$ of $Z$ momentum w.r.t. the beam axis
\item the helicity angle of each boson: $\theta_{W_1}^\ast$,$\theta_{W_2}^\ast$, $\theta_{Z}^\ast$.
\end{itemize}
Since the bosons are off-shell, there are no self-energy corrections
due to AQGC at NLO.  Such contributions only arise at NNLO via
loop-graphs including at least one AQGC-vertex.  So in the considered
order of perturbation theory there are no changes in the
energy-momentum-relations compared to the SM, hence off shell effects
do not lead to additional kinematical variables.

The cross section $\td \sigma^{\rm NLO}$ in NLO is of second order in
the coupling constants $\alpha_i$ since the processes (\ref{proc})
contain at most one vertex with AQGC, \begin{equation} \td
\sigma^{\rm NLO} = \td \sigma^{\rm SM} + \sum_i \td\sigma^{\rm
int}_i \alpha_i + \sum_{k\ell}\td\sigma^{\rm
ac}_{k\ell}\alpha_k\alpha_\ell.\end{equation} Here $\td\sigma^{\rm
SM}$ denotes the standard model contributions, $\td\sigma^{\rm int}$
the interference term linear in $\alpha$, and $\td\sigma^{\rm ac}$ the
quadratic term containing only AQG-vertices. The full information on
the dependence on anomalous gauge couplings is incorporated in at most
5 different event dependent parameters, viz.
\begin{equation}\label{eq:weight}
\frac{\td\sigma^{\rm NLO}}{\td\sigma^{\rm SM}} = 1 + A\alpha_4
+ B\alpha_4^2 + C\alpha_5 + D\alpha_5^2 + E\alpha_4\alpha_5,
\end{equation}
where the parameters $A,\ldots , E$ depend on the kinematical
variables, but not on anomalous couplings. They can be obtained by
re-weighting the SM matrix elements at five different points in the
$\alpha_4$-$\alpha_5$-plane for a fixed event sample and solving the
respective system of linear equations obtained from
Eq. (\ref{eq:weight}). We choose $\alpha_4 = \pm 0.1,\alpha_5 = 0$;
$\alpha_4=0, \alpha_5 = \pm 0.1$, and $\alpha_4=\alpha_5 = 0.1$.  The
resulting $\chi^2$ is an analytic function of $\alpha_i$ due to
(\ref{eq:weight}) and can easily be minimized.

\subsection{Particle Identification and Cuts}

All processes used in the analysis are listed in
Tab.~\ref{tab:6qstates}.  In the following, for convenience
and also historic reasons, processes that contain
intermediate $t\bar t$-production are  referred to as 
``background'', whereas those containing intermediate boson states
only are termed ``signal'' events.   In fact $kt$ production 
vastly dominates the triple boson production
 by two orders of magnitude (see Tab. \ref{tab:6qstates}). 
 Note that a negligible fraction of three boson production is
also included in the ``background'', since they cannot
(at generator level) be separated
from each other in a six quark production.

\begin{figure}[hbt]
\epsfig{file=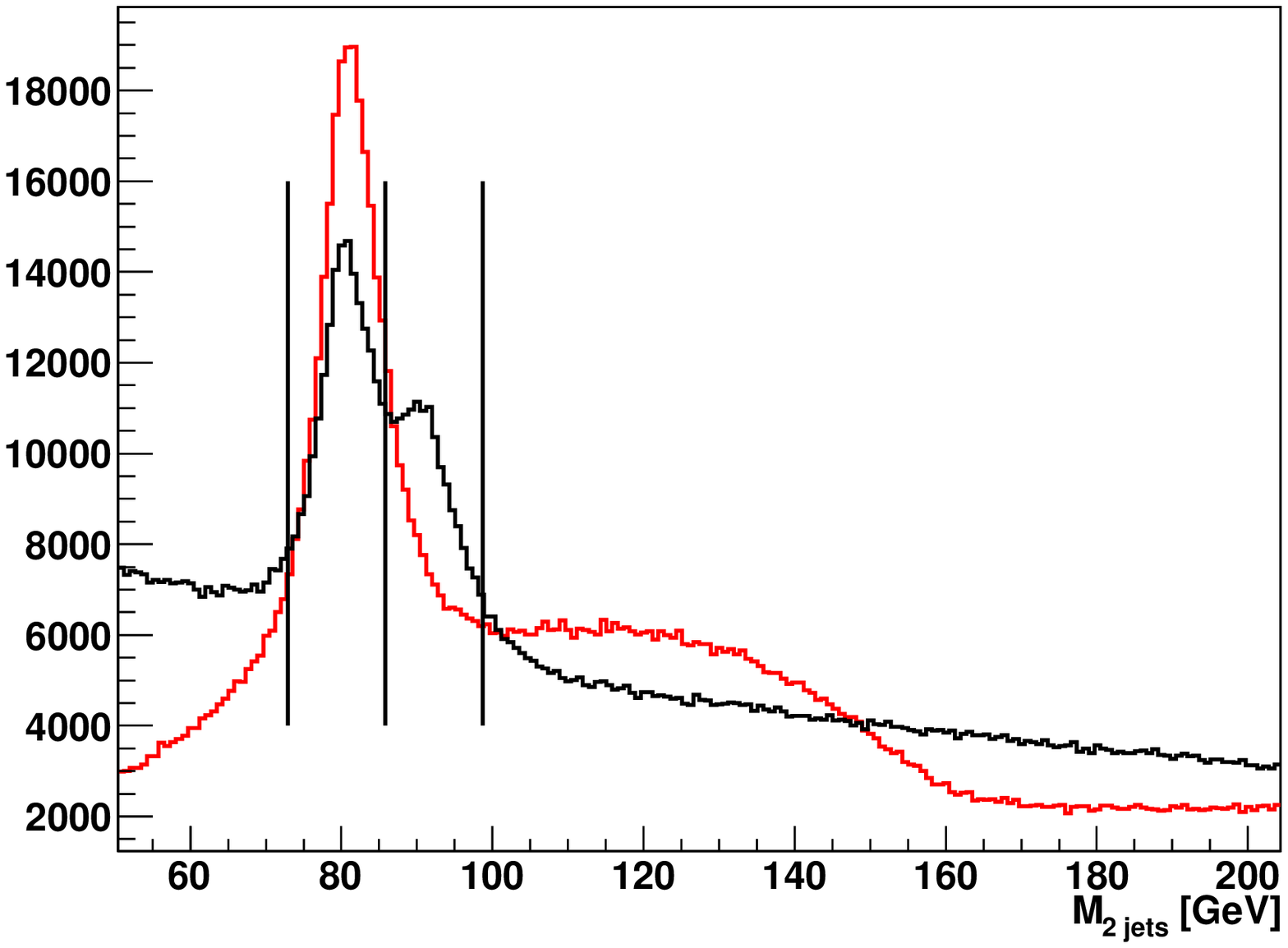,
width=0.45\textwidth}\\
\epsfig{file=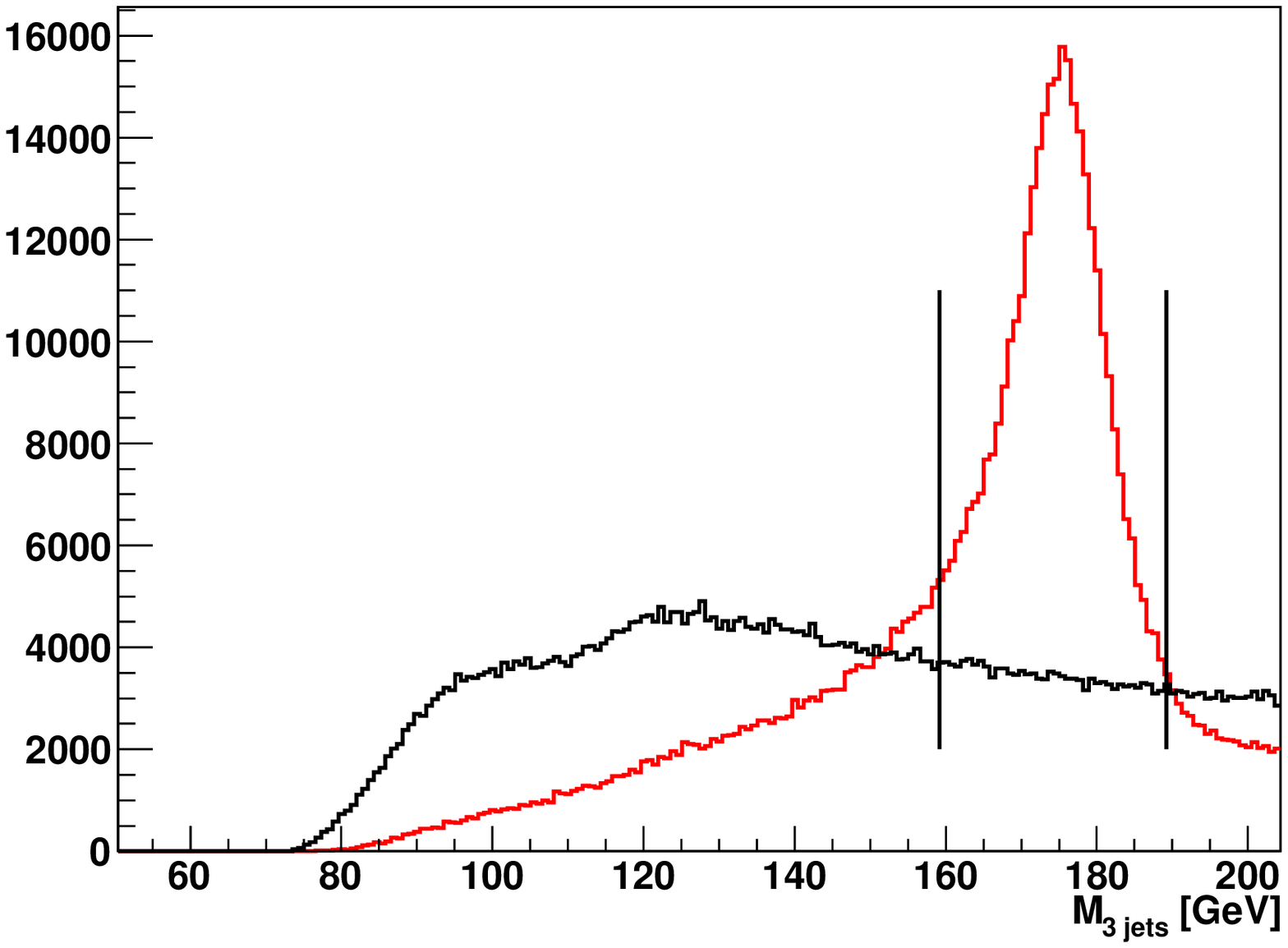,width=0.45\textwidth}\\
\caption{\label{fig:mass_cuts}The upper figure shows the invariant mass of jet pairs from which the bosons are reconstructed for background (red) 
and events without $tt$-production (black). The vertical lines
indicate the selection parameters $M_W = 72.9 {\ \rm GeV}\ldots 85.8{\
\rm GeV}$, $M_Z = 85.8{\ \rm GeV}\ldots 98.7{\ \rm GeV}$. Tops are
reconstructed by combining the best $W$ in the event with all
remaining jets and selected as indicated by the vertical lines in the
lower figure $|M_t - M_t^{\rm nom}| \leq 15 {\ \rm GeV}$.}
\end{figure}

Due to hadronization the six quark final states form jets.  Jet
identification is done using the Durham
algorithm~\cite{Pohl:2002vk}. In addition, we require a cut on the
jet energy $E_{\rm jet} > 5\ {\rm GeV}$ and an upper bound on
missing energy and momentum $E^2_{\rm mis} + p^2_{\perp} < (65\ {\rm
GeV})^2$ to accept only well defined jets.  The reconstructed jet
momenta are pairwise combined to form invariant masses $M_{{\rm jet
pairs}}= \sqrt{(p_{j1} + p_{j2})^2}$ of the boson candidates. Tops
are reconstructed via b-decay combining 3 jets and requiring two of
them to form a $W$.  Cuts on respective invariant masses used to
identify vector bosons and top quarks are given in
Tab.~\ref{tab:mass_cuts} and Fig.~\ref{fig:mass_cuts}
\begin{table}[htb]
\begin{tabular}{crr}
\hline\hline
particle & lower limit          & upper limit \\
\hline
$W$  &    $M_W - 7.5\ {\rm GeV}$    &   $(M_W + M_Z)/2$\\
$Z$  &    $(M_W+M_Z)/2$     &   $ M_Z + 7.5\ {\rm GeV}$\\
$t$  &    $ M_t - 15\ {\rm GeV}$    &   $M_t + 15\ {\rm GeV}$\\
\hline\hline
\end{tabular}
\caption{\label{tab:mass_cuts}Mass cuts used for particle reconstruction.}
\end{table}
Invariant masses of accepted candidates are required to have a minimum
deviation from the nominal masses $M_W = 80.403$ GeV, $M_Z=91.1876$
GeV, and $M_t = 174.2$ GeV as given in \cite{Yao:2006px}.

\begin{figure}[htb]
\subfigure[\label{fig:e_dist}Energy distribution of signal (black) and background (red).]{\epsfig{file=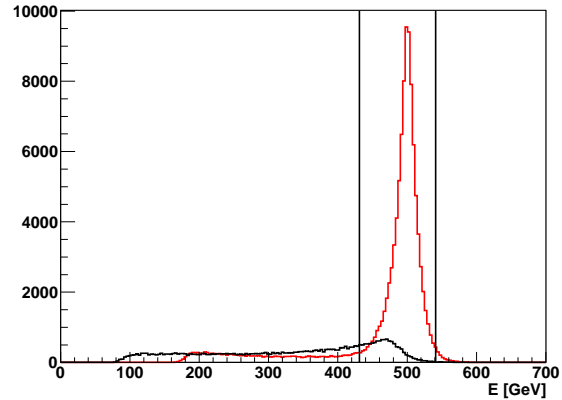, width=0.45\textwidth}}
\subfigure[\label{fig:p_dist}Momentum distribution of signal (black) and background (red).]{\epsfig{file=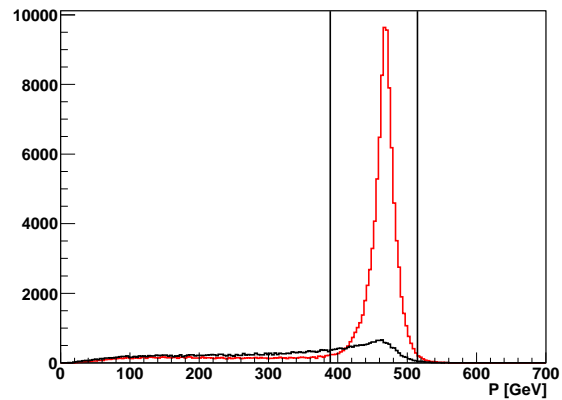, width=0.45\textwidth}}
\caption{\label{EP_cuts} The cuts applied to top candidates after detector simulation are shown:
$|E - 486\ {\rm GeV}| \leq 55\ {\rm GeV}$, $|p-452\ {\rm GeV}| \leq
63\ {\rm GeV}$. These numbers are mean values obtained from Monte
Carlo simulation of background events $e^+e^-\rightarrow t\bar t$. The
plots show (a) the $E$ and (b) the $p$ distribution from $tt$ 
background (red curve) and from the $WWZ$ signal (black curve).}
\end{figure}

For background identification we cut on the energy $E$ and momentum
$p$ distribution of the top candidates in the event as shown in
Fig.~\ref{EP_cuts}.
As ``signal'' events, only those are included that do not
contain any contributions from $t\bar t$-production, whereas the
``background'' consists to the largest part of such events. 

To enforce conservation laws a kinematical fit is applied, which
minimizes the missing energy and momentum constrained by the reconstructed
mass, i.e. the following expression for the reconstructed particles (reco) is minimized
\[ \sum_{\rm reco}\left\{ \frac{E^2_{\rm mis}}{\sigma^2_E} + \sum_{i=1}^3 \frac{P^2_{{\rm mis},i}}{\sigma^2_{P_i}} + \left(\frac{E_{\rm mis}^2 - P_{\rm mis}^2 - M^2_{\rm rec}}{\sigma_M^2}\right)^2\right\}\]
with respect to $\ E_{\rm mis}, \vec{p}_{\rm mis}$, where $E_{\rm mis}=\sqrt{s}-E_{\rm visible}$, $\vec{p}_{\rm mis} = -\vec{p}_{\rm visible}$.

\section{Sensitivities}

\begin{figure}[hbt]
\subfigure[\label{chi2wwz}$WWZ$]{\epsfig{file=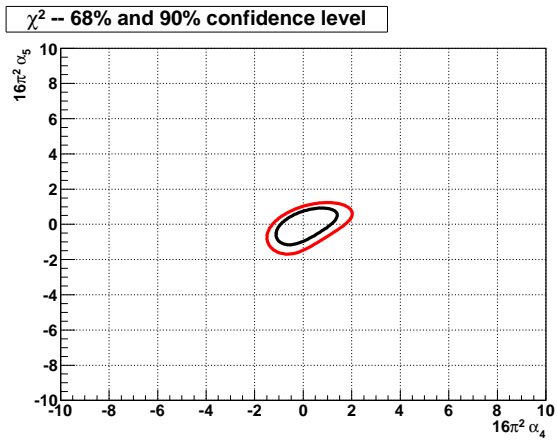,
width=0.45\textwidth}}\\
\subfigure[\label{chi2zzz}$ZZZ$]{\epsfig{file=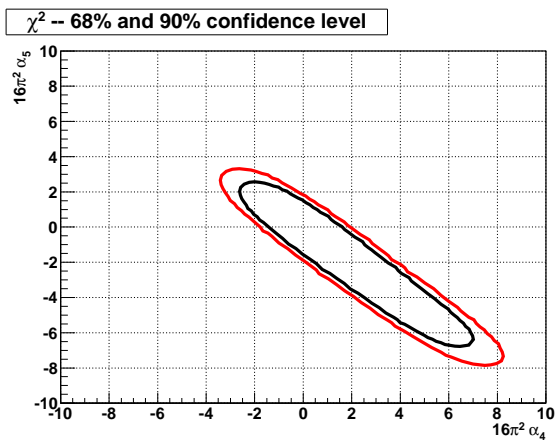,
width=0.45\textwidth}}
\caption{\label{chi2}Obtained sensitivity for reconstructed events.
Inner and outer contour delineate 68 \% and 90 \% confidence level,
respectively.}
\end{figure}
\begin{figure}\epsfig{file=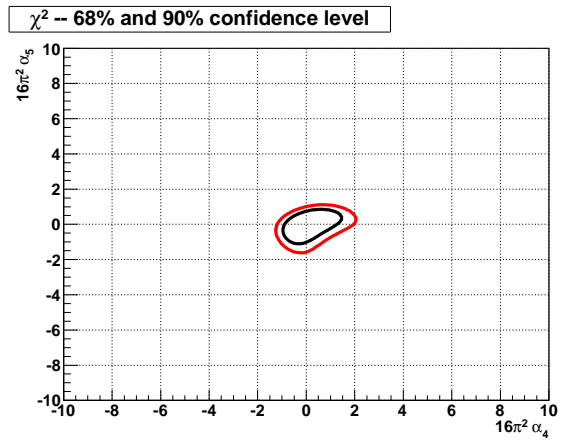,
width=0.45\textwidth}
\caption{\label{fig:chi_comb}Combined fit from reconstructed WWZ and ZZZ events.}
\end{figure}
 As a measure of sensitivity we use the 1-$\sigma$ standard
deviations of the $\alpha$ parameters obtained from minimizing
$\chi^2$ of Eq. (\ref{eq:chi2}) with a SM distribution as null
hypothesis.  Here, we implicitly assume that the standard deviations
(sensitivities) do not depend on the actual values of the anomalous
couplings. This is reasonable as long as the true values for the
anomalous couplings are unknown and are at least expected to lie
within the range where the effective theory approach is feasible.
The resulting contour gives an estimation of the maximum probability
that a measured event distribution is compatible with the null
hypothesis.  The sensitivities obtained by this procedure are shown
in Fig.
\ref{chi2} for intermediate WWZ and intermediate ZZZ events,
respectively. 
The combined $\chi^2$ fit using reconstructed $WWZ$ and $ZZZ$ events
can be found in Fig.~\ref{fig:chi_comb}.  Tab.~\ref{qagc} shows the
1-$\sigma$ standard deviation for the coupling parameters calculated
from Minuit.

\begin{table}[h]
\begin{tabular}{cc@{\ \ }ccccc}
\hline\hline
& & \multicolumn{2}{c}{Ref. \cite{Beyer:2006hx}} & \multicolumn{3}{c}{this work, 6 quarks}  \\
   &    &               & WWZ   & (WWZ)  & (ZZZ) & combined \\
\hline
$16\pi^2\alpha_4$ & $\sigma^-$ && -3.53 &  -1.49  & -1.87 & -0.68 \\
       & $\sigma^+$  && 1.90 &  2.02  &  5.08&  0.95 \\
\hline
$16\pi^2 \alpha_5$ & $\sigma^-$  &&-1.64  &  -1.69  & -5.20&  -0.71\\
       & $\sigma^+$   & &3.94 &  1.24  &  1.85&   0.60\\
\hline
\end{tabular}
\caption{\label{qagc}1-$\sigma$ standard deviations for $\alpha_4$ and $\alpha_5$ obtained from reconstructed
$WWZ$, $ZZZ$ events and from a combined fit. The lower table shows the
previously obtained results (taken from \cite{Beyer:2006hx}).}
\end{table}

\clearpage
\section{Conclusion}

Based on our previous analysis the sensitivity on quartic gauge
couplings of processes at the ILC involving three intermediate bosons
have been improved. The improvement has been in two directions: First
we have included observables related to the polarization of the
intermediate bosons.  This leads to a better sensitivity compared to
the previous results.  Second, we have utilized full six quark final
states in the event generation that became available only
recently. Matrix elements of all six fermion final states that can be
accessed in the 1~TeV range have been included. This leads to a more
realistic description of the processes closer to the expected
experimental behaviour. It also leads to a more realistic treatment of
background events compared to former analysis where only
non-interfering $tt$-production has been assumed.

Of all possible generic quartic gauge couplings the production of
intermediate three boson states $WWZ$ and $ZZZ$ is sensitive to the
ones involving $\alpha_4$/$\alpha_6$, and $\alpha_5$/$\alpha_7$. 
The present analysis
is done for the ILC with the option for initial state polarization. Due to
the large number of events that have to be considered, we have merely
favored a scenario of fully polarized initial states that has been the
most sensitive case in our previous analysis~\cite{Beyer:2006hx}

Presently, we have only utilized hadronic decays. These amount to 32\%
of the total events with intermediate three bosons.  A more complete
investigation including semileptonic decays, i.e.  $e^+e^-\rightarrow
(VVZ) \rightarrow 4q + \ell\bar\ell$ is left for future work.

A next task is to combine the present analysis with the results from
$WW$-scattering along the lines of~\cite{Beyer:2006hx} and update the
characteristics of new resonant states that can be inferred from the
present work.

\bibliographystyle{unsrt}
\bibliography{EWSB}

\begin{thebibliography}{10}

\bibitem{Susskind:1978ms}
Leonard Susskind.
\newblock {Dynamics of Spontaneous Symmetry Breaking in the Weinberg- Salam
  Theory}.
\newblock {\em Phys. Rev.}, D20:2619--2625, 1979.

\bibitem{Kilian:2003pc}
W.~Kilian.
\newblock {Electroweak symmetry breaking: The bottom-up approach}.
\newblock {\em Springer Tracts Mod. Phys.}, 198:1--113, 2003.

\bibitem{Longhitano:1980tm}
Anthony~C. Longhitano.
\newblock {Low-Energy Impact of a Heavy Higgs Boson Sector}.
\newblock {\em Nucl. Phys.}, B188:118, 1981.

\bibitem{Krstonosic:2005qp}
P.~Krstonosic, K.~Monig, M.~Beyer, E.~Schmidt, and H.~Schroder.
\newblock {Experimental studies of strong electroweak symmetry breaking in
  gauge boson scattering and three gauge boson production}.
\newblock 2005.

\bibitem{Beyer:2006hx}
M.~Beyer et~al.
\newblock {Determination of new electroweak parameters at the ILC: Sensitivity
  to new physics}.
\newblock {\em Eur. Phys. J.}, C48:353--388, 2006.

\bibitem{Kilian:2005bz}
W.~Kilian and J.~Reuter.
\newblock {Resonances and electroweak observables at the ILC}.
\newblock 2005.

\bibitem{Moretti:2001zz}
Mauro Moretti, Thorsten Ohl, and Jurgen Reuter.
\newblock {O'Mega: An optimizing matrix element generator}.
\newblock 2001.

\bibitem{Kilian:2001qz}
W.~Kilian.
\newblock {WHIZARD 1.0: A generic Monte-Carlo integration and event generation
  package for multi-particle processes. Manual}.
\newblock LC-TOOL-2001-039.

\bibitem{Sjostrand:2003wg}
Torbjorn Sjostrand, Leif Lonnblad, Stephen Mrenna, and Peter Skands.
\newblock {PYTHIA 6.3: Physics and manual}.
\newblock 2003.

\bibitem{Pohl:2002vk}
M.~Pohl and H.~J. Schreiber.
\newblock {SIMDET - Version 4: A parametric Monte Carlo for a TESLA detector}.
\newblock 2002.

\bibitem{James:1975dr}
F.~James and M.~Roos.
\newblock {Minuit: A System for Function Minimization and Analysis of the
  Parameter Errors and Correlations}.
\newblock {\em Comput. Phys. Commun.}, 10:343--367, 1975.

\bibitem{Yao:2006px}
W.~M. Yao et~al.
\newblock {Review of particle physics}.
\newblock {\em J. Phys.}, G33:1--1232, 2006.

\end{thebibliography}

\end{document}